# Quantitative implications of the secondary role of carbon dioxide climate forcing in the past glacial-interglacial cycles for the likely future climatic impacts of anthropogenic greenhouse-gas forcings

Willie Soon (wsoon@cfa.harvard.edu)
Harvard-Smithsonian Center for Astrophysics, Cambridge, Massachusetts 02138, USA

**Abstract:** A review of the recent refereed literature fails to confirm quantitatively that carbon dioxide ($CO_2$) radiative forcing was the prime mover in the changes in temperature, ice-sheet volume, and related climatic variables in the glacial and interglacial periods of the past 650,000 years, even under the "fast response" framework where the convenient if artificial distinction between forcing and feedback is assumed. Atmospheric $CO_2$ variations generally follow changes in temperature and other climatic variables rather than preceding them. Likewise, there is no confirmation of the often-posited significant supporting role of methane ($CH_4$) forcing, which – despite its faster atmospheric response time – is simply too small, amounting to less than 0.2 W/m$^2$ from a change of 400 ppb. We cannot quantitatively validate the numerous qualitative suggestions that the $CO_2$ and $CH_4$ forcings that occurred in response to the Milankovich orbital cycles accounted for more than half of the amplitude of the changes in the glacial/interglacial cycles of global temperature, sea level, and ice volume. Consequently, we infer that natural climatic variability – notably the persistence of insolation forcing at key seasons and geographical locations, taken with closely-related thermal, hydrological, and cryospheric changes (such as the water vapor, cloud, and ice-albedo feedbacks) – suffices *in se* to explain the proxy-derived, global and regional, climatic and environmental phase-transitions in the paleoclimate. If so, it may be appropriate to place anthropogenic greenhouse-gas emissions in context by separating their medium-term climatic impacts from those of a host of natural forcings and feedbacks that may, as in paleoclimatological times, prove just as significant.

## 1. A rare but incomplete consensus on the relationship between $CO_2$ concentration, temperature, and ice-sheet volume

One of the most notable, but somewhat surprising, consensus conclusions from ice-core drilling projects and researches at both poles (see e.g., Fischer et al. 2006; Masson-Delmotte et al. 2006) is the fact that the deduced isotopic temperatures lead other climatic responses, including especially the atmospheric levels of minor greenhouse gases like $CO_2$ and $CH_4$. Fischer et al. (1999) first reported that atmospheric $CO_2$ concentrations increased by 80 to 100 ppm 600± 400 years after the warming of the last three deglaciations (or glacial terminations) in Antarctica and that relatively high $CO_2$ levels can be sustained for thousand of years during glacial inception scenarios when the Antarctic temperature has dropped significantly. Later, Monnin et al. (2001) and Caillon et al. (2003) offered clear evidence that temperature change drove atmospheric $CO_2$ responses during more-accurately dated periods near glacial terminations I (at about 18 kyr before present, BP) and III (at about 240 kyr BP), respectively.

Stenni et al. (2001) presented evidence that the rising tendency of atmospheric $CO_2$ can be interrupted by abrupt events like the Antarctic Cold Reversal and related Oceanic Cold Reversal starting around 14 kyr BP, so that factors like the changes in the southern ocean, high-latitude atmospheric circulation and dust transport can quickly trigger responses in global carbon cycling. Figure 1 offers the graphical summary of the apparent consensus on atmospheric $CO_2$ content being driven by change in temperature using the Vostok data first published by Petit et al. (1999). The latest 650 kyr-long record from the EPICA Dome Concordia effort reported by Siegenthaler et al. (2005) remarkably confirmed the $CO_2$-temperature lag relation over terminations V, VI and VII, covering the period from 400 to 650 kyr BP. The analyses by Delmotte et al. (2004) also confirmed that atmospheric $CH_4$ systematically lags the proxy for Antarctic temperatures by 1100±200 years, adopting the long timescale windows of 50 to 400 kyr for the lead-lag analyses.

Additionally, there are new and innovative analyses of both high-resolution ice and gas data from Antarctica and Greenland (Loulergue et al. 2007; Ahn and Brook 2007; Kobashi et al. 2007) that are able to better time the complicated relation of atmospheric $CO_2$ and $CH_4$ with local and regional temperatures at both locations for specific abrupt warming and cooling events like the Dansgaard-Oeschger warming, Heinrich icebergs-rafting or even the 8.2 kyr cooling events in Greenland. Ultimately, certain local and regional climatic variations and changes must be responsible for driving the responses of the global carbon budget cycles—otherwise, some extra-terrestrial factors and/or extra-ordinary scenarios would need to be invoked to explain the trigger and maintainence of the atmospheric $CO_2$ and $CH_4$ variations every 100 kyr or so for the past 650 kyr. Such a picture is not inconsistent with the recent comprehensive review by Peacock et al. (2006) that manages to specify a combination of distinct climatic, oceanic and bio-chemical processes that could explain the co-variation of atmospheric $CO_2$ during a 100 kyr glacial-interglacial transition.

Synthesis studies by Shackleton (2000), Mudelsee (2001), Pepin et al. (2001), and Ruddiman and Raymo (2003) pointed out that both Antarctic temperatures and atmospheric $CO_2$ concentrations significantly lead the changes in the large ice-sheets of the Northern Hemisphere by at least a few thousand years to as long as 14 kyr over the full 420 kyr of the Vostok record. The scenario that seems most reasonable is that the external orbital insolation forcing triggered the fast and large changes in the air temperature in the Antarctic/Southern Hemisphere which, in turn, caused responses in deep-ocean properties (including changes in both the temperature and the volume of southern-sourced deep-water filling the ocean basins; see Skinner 2006) and in the global carbon cycle leading to the changing levels of atmospheric $CO_2$, which ultimately acted as an amplifier for the glacial-interglacial ice volume variability. Several recent pollen, glacial, and reflectance (i.e., natural gamma rays) records from the mid-latitude Southern Hemisphere clearly support such a scenario with insolation changes in the southern-hemisphere first and then followed by related thermal, chemical, hydrologic and cryospheric responses (Carter and Gammon 2004; Suggate and Almond 2005; Vandergoes et al. 2005; Sutherland et al. 2007).

Alley et al. (2002) and Johnston and Alley (2006) offered a different scenario by emphasizing a northern-hemisphere thermal lead or trigger (see also the new evidence and discussion for the past 360 kyr in Kawamura et al. 2007), instead of a southern-lead scenario, for the consequential chains of variations in solar insolation, air and sea temperatures, $CO_2$ and global ice volume (mainly of the bulk continental ice sheet in the Northern Hemisphere), but this alternate picture clearly also proposes atmospheric $CO_2$ as the essential amplifier of the warming and cooling and hence the waxing and wanning of global ice volume.

## 2. Atmospheric $CO_2$ radiative forcing as an amplifier of glacial-interglacial changes and its other theoretical roles

It is still unclear as to whether such a plausible $CO_2$-amplification scenario can be quantitatively confirmed with available evidence to-date. The very long lead time by the radiative forcing of atmospheric $CO_2$ requires better clarification through physical modeling to account for the full dynamics of 1) the great continental ice sheets, i.e., which can sustained for 10 kyrs in the phase of maximum ice extent like that deduced for the Laurentide Ice Sheet during the Last Glacial Maximum, LGM (Dyke et al. 2002), or for the extension of the ice sheet over the Barents and Kara Seas into the continent (Svendsen et al. 1999), and of 2) the ice conditions around the Arctic Ocean (Norgaard-Pedersen et al. 2003; Martinson and Pitmann 2007) and their interactions with the coupled dynamics of the ocean and atmosphere, especially at the tropics (see e.g., Ashkenazy and Tziperman 2006; Roe 2006).

It is also difficult to presume any significant amplifying role by atmospheric $CO_2$ for the extremely large winter cooling of 28ºC over Greenland-Arctic-North Atlantic area during the Younger Dryas event recently hypothesized by Broecker (2006). Similarly Barrows et al. (2007) noted several periods of rapid SST changes in their southern midlatitude records around 58-38 kyr BP (i.e., Oxygen Isotope Chronozone or Marine Isotope Stage, MIS, 3) and around 19.5-18.5 kyr BP (i.e., where a rapid warming of 5ºC in 130 years was recorded at 19.4 kyr BP and a similarly large and rapid cooling starting around 19.2 kyr BP) that are simply difficult to be explained by atmospheric $CO_2$ forcing.

More importantly, Roe (2006) showed that if one adopts the rate of change of global ice volume rather than the absolute ice volume as the physically more direct measures of ice-sheet dynamics and its thermal connections to summertime temperature and local summertime insolation radiation, then the sketched scenario of changing $CO_2$ leading the change in the global ice volume or ice sheet may not be correct. Instead, careful analyses by Roe (2006) found that atmospheric $CO_2$ actually lags the rate of global ice volume change by a few thousand years, or $CO_2$ is at most synchronous with the rate of change in global ice volume. Roe (2006) concluded that "... variations in melting precede variations in $CO_2$ ... This implies only a secondary role for $CO_2$ — variations [which] produce a weaker radiative forcing than the orbitally-induced changes in summertime insolation [discussed in section 3 below] — in driving changes in global ice volume." These issues will be discussed further in sections 3 and 4.

Another important empirical consensus that has recently emerged is the fact that older estimates of no-more-than 1-2ºC changes in the tropical sea surface temperatures (SST) during the Last Glacial Maximum (LGM) of about 21 kyr BP, or during other glacial-interglacial transition periods, may have been seriously underestimated (e.g., Schrag et al. 1996; Peltier and Solheim 2004). Lea et al. (2000) showed that the equatorial SSTs in the core area of the western Pacific warm pool, a region that could be sensitive indicator of the global carbon cycling, were 2.8º ± 0.7 ºC colder at the LGM than at present. Visser et al. (2003) recently suggested that SST around the Indo-Pacific warm pool area by the Makasar strait varied by about 3.5-4ºC during the last two glacial-interglacial transitions. Barrows and Juggins (2005) synthesized the proxy SST records from about 165 cores for oceans around Australia, including the Indian Ocean and their compilations showed a cooling of up to 4ºC in the tropical eastern Indian Ocean and up to 7 to 9ºC in the higher latitude regions of the southwest Pacific Ocean during the LGM. Three new alkenone-derived SST records from the midlatitude southern hemisphere presented by Pahnke and Sachs (2006) essentially confirmed the large amplitude change in SST between glacial and interglacial transitions. The consequence of that large amplitude SST change in the tropics and higher latitudes has led Visser et al. (2003) to suggest that a substantial portion of the 80 ppm change in atmospheric $CO_2$ during a glacial-interglacial transition can be simply explained by a direct change in $CO_2$ solubility in sea water as a function of SST change (see Peacock et al., 2006, for additional processes involving changes in sea level, oceanic circulation and related chemical and biological responses). Peltier and Solheim (2004) offer a quantitative estimate in explaining "more than 60%" of the observed changes in glacial-interglacial $CO_2$ contents from air bubbles trapped in ice cores.

There are two additional roles often assigned to the $CO_2$ radiative forcing that highlight the remarkable sensitivity of the climate system behavior and evolution to atmospheric $CO_2$ (e.g., Saltzman et al. 1993). We will only summarize these briefly here.

First, there are theoretical speculations, on even longer geologic timescales, about the formation of large icesheets in both hemispheres since about 2.7 million years BP (or late Pliocene transition, see e.g., the overviews in Droxler and Farrell 2000; Crowley and Berner 2001; Lisiecki and Raymo 2007) and about the transition from the "41 kyr world" to the "100 kyr world" (i.e., actually more like ice age cycles every 80 kyr to 120 kyr or so; see Liu and Herbert 2004; Huybers and Wunsch 2005) starting around 780 kyr ago or in the mid-Pleistocene transition, or MPT (see Berger et al. 1999). (Clark et al. 2006 described the MPT as a broader transitional variability zone from 1250 kyr to 700 kyr BP.) It was suggested that the two phenomena/events happened because the Earth's climate system has been undergoing gradual global cooling from a systematic decrease in atmospheric $CO_2$ until a dynamical air-sea-ice sheet interaction threshold is crossed. However, the evidence for an overall cooling since 780 kyr BP is not that strong, and the idea has been recently challenged by de Garidel-Thoron et al. (2005). Furthermore, non-$CO_2$-related explanations involving changes in basal conditions under ice sheets (Clark et al. 1999) may be significant, and that the MPT may simply be a dynamical system response to the continuous obliquity pacing as shown and discussed by Liu and Herbert (2004), Huybers and Wunsch (2005), and Huybers (2007). As for the explanation of the late Pliocene cooling, Revelo et al. (2004) argued that the transition from warm early-to-mid Pliocene into increasing glaciation in the northern hemisphere is linked more to nonlinear coupled dynamics of ocean and atmosphere that altered the meridional heat and moisture transfers than to any persistent cooling or threshold-crossing tendency due to a decrease in atmospheric $CO_2$ over the past 3 million years. In two related modeling studies, Barreiro et al. (2006) and Fedorov et al. (2006) found that the physical explanation for the great warmth during the early-to-middle Pliocene from 5 to 3 million years BP may be connected to the reality of a permanent, rather than intermittent, El Nino condition (see, however, the challenge by Haywood et al. 2007) plausibly involving the collapse of trade winds along the equator with the attendant large decrease in low-level stratus clouds, hence a large increase in incoming solar radiation and increase in the atmospheric water vapor feedback in heating up the tropical atmosphere and ocean. Huybers and Molnar (2007) recently concluded that a long-term cooling trend in the eastern tropical Pacific alone (which they clearly distinguished this explanation from the one specifying high $CO_2$ in the early-to-mid Pliocene), rather than a permanent El Nino scenario, may be sufficient for explaining the increased glaciation since 3 million years BP.

The second prominent effect by $CO_2$ radiative forcing has been framed as follows. Two new studies are now suggesting that the observed variation of about 30 ppm in the $CO_2$ concentrations during interglacial periods may contribute significantly to thermal and moisture instabilities that eventually drove glacial advances (Vettoretti and Peltier 2004; Kubatzki et al 2006). But the empirical basis for arguing about such a strong non-linear effect of $CO_2$ forcing on the climate system evolution and change is not strong considering the small amplitude of the $CO_2$ radiative forcing (see also comments on p. 275 in Khodri et al. 2003), the $CO_2$-temperature lagged-response relation, and even the actual simulated results (i.e., the rather small relative differences in the near-surface global temperature of no more than 0.5ºC and inland ice-sheet area of no more than 0.5 million $km^2$ over North America shown in Figure 9 of Kubatzki et al. 2006 for two different $CO_2$ radiative forcing scenarios). In contrast, Bauch and Kandiano (2007) point to significant differences in the surface ocean conditions during the previous interglacial and the Holocene while supporting the interpretation of the significant variabilities on centennial and millennial timescales during interglacials as related to intrinsic variations in solar irradiance outputs and subsequent amplification through mechanisms like solar effects on distribution and transport of Arctic sea ice and changes in the meridional overturning circulation of the North Atlantic as proposed earlier by Bond et al. (2001) for the Holocene.

The next two sections (3 and 4) focus more narrowly on the inability to find quantitative support for the putative role of $CO_2$ radiative forcing in the observed glacial-interglacial cycles of global ice volume and temperatures over the past 650 kyr as dictated by the latest EPICA Dome Concordia's records of climatic variability and trace-gases history (EPICA community members 2004; Siegenthaler et al. 2005; Spahni et al. 2005). Conclusions are given in section 5.

**3. Explaining glacial-interglacial climate change and environmental responses by orbitally-moderated insolation forcing acting locally and regionally compared with the effects of global radiative forcing from changing $CO_2$ concentration**

Soon et al. (2001, p. 261) earlier cautioned against the premature rejection of the role of changes in solar radiation (i.e., from both orbital motion-induced and intrinsic solar magnetism-caused variability) in favor of the rather simplistic, and very possibly incorrect, picture of the domination of the climate system by changes in the global radiative forcing related to the man-made emission of atmospheric greenhouse gases like $CO_2$ and $CH_4$. The caution was partly related to the illustrative and yet successful modeling experiments showing the nonlinear dynamical responses of ice sheets to the lone modulation in seasonality of solar insolation by Posmentier (1994).

Figure 2 shows the sharp contrast in the amplitude of variability between global annual-mean insolation and daily summer insolation at 65ºN (Laskar et al. 1993). It shows the remarkably small changes of no larger than 0.6 W/m$^2$ in the net global solar radiation induced by the orbital evolution of the Earth around the Sun over the past 1 million years. This estimate, however, has not accounted for intrinsic changes to the Sun's irradiance as modulated by solar magnetic activity; an amplitude change of a few W/m$^2$ over a million years cannot be ruled out. But Figure 2 confirms that the large changes in the summer solar insolation at the key ice-nucleation location like 65ºN (see Roe 2006) are clearly much larger than the relatively small global radiative forcing by changes of $CO_2$ for the glacial-interglacial transitions of the past 650 kyr, estimated to be 2-3 W/m$^2$. From the perspective of the climatic system, local summer insolation, as long as the persistency is guaranteed by the locked-in orbital motion, is a highly relevant physical quantity for studying and quantifying the local and regional responses of the thermal and hydrologic variables, and may ultimately explain the apparent global statistics of regional weather and climate. Vandergoes et al. (2005) and Sutherland et al. (2007) have emphasized the important role of local summer solar insolation forcing rather than any global annual-mean measure with the view of southern-hemisphere-lead scenario discussed in section 1 above. Lorenz et al. (2006) supported the key role of local and regional insolation changes and emphasized the nonlinear changes in the entire seasonal cycle of insolation for the spatial heterogeneity of the Holocene climate trends. Empirical support for this view comes from the apparent dominance of the global radiative forcing of 2 W/m$^2$ (e.g., Joos 2005) estimated from the 80 ppm change in atmospheric $CO_2$ over the glacial-interglacial cycles of the past 650 kyrs, and yet no clear climatic response can be confidently or completely linked to $CO_2$ as discussed here and in the following section.

A diametrically opposite conclusion has been reached by Archer and Ganopolski (2005) arguing for the great receptivity of the climate system to global radiative forcing by $CO_2$ when contrasted with the orbital insolation forcing. Their appeal to $CO_2$ forcing modulation of the threshold of ice age cycles is not dissimilar to some of the dramatic, but qualitative, ideas reviewed in section 2 above. This major disagreement should motivate further serious scientific inquiry. In the present brief paper concerning the quantitative role of $CO_2$ in glacial-interglacial changes it is necessary to postpone a more complete synthesis of the external solar-and-interstellar forcing in accounting for 1) intrinsic variability of the Sun from its thermo-nuclear and magnetic history (Gough 1990; Gough 2002; Turck-Chieze et al. 2005; and with important insights from geological archives e.g., Sharma 2002; Lal et al. 2005; Bard and Frank 2006), 2) both the long-term (Berger 1978; Laskar et al. 2004) and shorter term (i.e., from multi-years to decades to centuries; Loutre et al. 1992) perturbations of the Earth's orbital geometry with respect to the Sun, and (3) even for any intrinsic variability related to the local interstellar (Frisch and Slavin 2006; Muller et al. 2006; Scherer et al. 2006) and galactic (Cox and Loeb 2007) environments.

One can however note that most constructions of physical theory and modeling of glacial and interglacial changes (Kukla and Gavin 2005; Roe 2006; Tziperman et al. 2006; Huybers 2007; Martinson and Pitman 2007) do not require $CO_2$ to be a predominant forcing agent but instead strongly hint at both the necessary and sufficient conditions of orbital insolation forcing, its persistency and its pacing role through nonlinear phase locking. A direct comparison of the 80 ppm change in atmospheric $CO_2$ for a radiative forcing of about 2 $W/m^2$ (e.g., Joos 2005) with the 10 $W/m^2$ summertime shortwave forcing, after properly folding in the albedo of melting ice and summer half-year insolation variation by Roe (2006), provide us with a clear hint about the secondary role of $CO_2$ in setting the trend in climate change and other related responses during the glacial-interglacial transitions of the past 650 kyr. The estimate for the radiative forcing of 0.2 $W/m^2$ (e.g., Joos 2005) by atmospheric $CH_4$ change of about 400 ppb over the 100-kyr glacial-interglacial cycle also does not suggest a very prominent role by $CH_4$ either in isolation or in combination with $CO_2$.

At this stage, it may be also relevant to point out that the popular scenario for potential episodic releases of methane hydrates to act as a strong positive feedback commonly tied to seed atmospheric warming by $CO_2$ may not be so straightforward. First, Milkov (2004) has cautiously lowered the previously accepted high-estimate of global hydrate-bound gas from 21 x $10^{15}$ $m^3$ of methane (or about 10,000 Gt of methane carbon) to a much lower range between 1 to 5 x $10^{15}$ $m^3$ of methane (or about 500-2500 Gt of methane carbon). Next, Cannariato and Stott (2005) have recently challenged the possibly incorrect interpretation of the large $\delta^{13}C$ excursions in records of planktonic and benthic foraminifera as clathrate-derived methane release. A careful examination of the atmospheric methane carbon isotope ratio ($\delta^{13}CH_4$) from western Greenland ice margin spanning the Younger Dryas-to-Preboreal transition by Schaefer et al. (2006) also could not find support for either catastrophic or gradual marine clathrate emissions. Finally, Bhaumik and Gupta (2007) have recently identified 5 major episodes of methane releases starting since 3.6 million years BP in their ODP 997A site located on the crest of the Blake Outer Ridge (about 200 km off the east coast of the United States from the shores of Georgia and South Carolina) to be probably linked to reduced hydrostatic pressure connected to lowered sea levels and intense glacial events roughly coinciding with increased glaciation in the northern hemisphere.

The bottom line is still that many numerical attempts (shown below) to quantify the impacts from variations in these two minor greenhouse gases simply do not confirm their predominant roles in explaining the large amplitude changes in thermal, hydrologic and cryospheric history during the glacial-interglacial transitions. The failure in quantitatively linking the seed $CO_2$-induced thermal perturbations to large hydrologic and cryospheric responses is the necessary reason for questioning the $CO_2$-amplifier idea. The persistent solar insolation forcing at key seasons and geographical locations and closely related thermal, hydrological and cryospheric changes (including the water vapor, cloud and ice-albedo feedbacks) may be sufficient to explain the regional and global climatic changes during the glacial-interglacial transitions.

**4. Computer simulations of climate provide little quantitative support for the hypothesis that $CO_2$ is a significant amplifier of global mean temperature**

Why is the climatic role of $CO_2$ radiative forcing deemed so hard to confirm?

Figure 3 may help explains the inherent difficulty in confirming any radiative impact from added $CO_2$ forcing using the deduced global net longwave (LW) fluxes available from the International Satellite Cloud Climatology Project (ISCCP) over the 18-year span from July 1983 through June 2001 despite some known data limitations for ISCCP (e.g., Kato et al. 2006; Evan et al. 2007). Over that period, the $CO_2$ increase is estimated to produce an equivalent global LW forcing of only about 0.3 W/m$^2$ and this amount is practically not discernible from the large interannual variability of LW fluxes at either the surface, the atmospheric air column, or even the top of the atmosphere. It is understood in scientific discussions, but popularly least appreciated, that the argument for a significant role from added radiative forcing by anthropogenic emissions of $CO_2$ rests on the assumption that over a sufficiently long interval, say over several decades to century, the Earth's climate system will be in some form of "equilibrium" state where all the intrinsic LW flux fluctuations shown in Figure 3 will cancel almost exactly to zero. The cancellation would in turn allow the detection and hence all related climatic manifestations, e.g., of a systematic increase of about 4 W/m$^2$ of net radiative LW forcing (or ranging from 3.5 to 4.2 W/m$^2$ in the $CO_2$ forcing parameterization of 20 GCMs examined by Forster and Taylor 2006) from the doubling of $CO_2$ content roughly over 70 year's time (i.e., a compounded rate of $CO_2$ increase at 1% per year).

It is thus widely accepted that although atmospheric $CO_2$ and $CH_4$ contents during the glacial-interglacial cycle is a response to climate-induced perturbations to the global carbon and methane reservoirs both in the land, surface ocean, continental margin and deep sea, the $CO_2$ and $CH_4$ radiative forcing can in turn act to strongly amplify and synchronize climatic changes across all weather regimes and climatic zones from south to north poles, ultimately producing a global warming or cooling. However, a closer look reveals that most of the claims, even in many scientific publications (i.e., from Genthon et al. 1987; Lorius et al. 1990; through Hansen et al. 2007), have not offered reliable quantitative supports for the claim. The necessity of apriori forward calculations for the difficult task of quantifying climatic role of $CO_2$ radiative forcing must be contrasted with multi-variable regression analyses as performed in those cited studies, adopting rather selective variables that can easily be confused by the concepts of forcing and feedback (see further discussion below). This may be the reason for the early caution issued by Genthon et al. (1987) that "$CO_2$ changes might just be a consequence of climatic change without much effect on climatic change itself."

Furthermore, there are clearly too many adjustable values in the estimates of radiative forcings by various other factors. For example, the aerosol-dust forcing during the LGM was estimated to be -1.0± 0.5 W/m$^2$ by Hansen et al. (1993) and then later modified to -0.5±1.0 W/m$^2$ in Hansen et al. (1997) (based on the modeling study of Overpeck et al. 1996, that has since been questioned by Claquin et al. 2003). A significantly larger estimate by Claquin et al. (2003) gave a global dust forcing during LGM that ranges from -1.0 to -3.2 W/m$^2$ with forcing at high-latitudes (poleward of 45º) from -0.9 to +0.2 W/m$^2$ and 15ºN-15ºS tropical forcing ranging from -2.2 to -3.2 W/m$^2$.

It is puzzling that well-accepted roles of atmospheric water vapor and cloud feedbacks (despite the fact that both variations in the isotopic proxies $\delta D_{ice}$ and $\delta^{18}O_{ice}$ from ice cores are essentially markers of large hydrologic changes) are not often factored-in or discussed more seriously when $CO_2$ as the amplifier of glacial-interglacial warming or cooling is being considered. This is especially so because varying levels of atmospheric $CO_2$ largely seem to be a climatic feedback response, rather than any external forcing as envisioned in the scenario of anthropogenic emissions of $CO_2$.[1] Other potentially powerful hydrologic feedbacks like the weakened hydrologic cycle during LGM from a significant lowering of the mean residence time of water vapor from excessive dust loading in the atmosphere proposed by Yung et al. (1996) also have not gained much attention compared to those from added radiative forcing by $CO_2$. Paleoclimatic studies discussing the modulation of greenhouse effects by water vapor and cloud formation, e.g., over warm ocean areas and follow-up effects by orbital forcing (e.g., Gupta et al. 1996) should be a priority. Priem (1997) even went as far as suggesting that the powerful greenhouse effects by water vapor, rather than $CO_2$, in the early Earth (up to one and a half billion years old) that consists mainly of oceans with little dry land could come a long way toward in resolving the "faint young Sun paradox".

Another example of potentially important feedback concerns how the seasonal cycle of surface-penetrating solar radiation is coupled to oceanic biota and the related biogeochemical emissions and atmospheric responses, especially through the production of marine biogenic dimethlysulfide (e.g., Shell et al. 2003; Vallina and Simo 2007). Finally, beneficial insights may also be gained from studying how orbital forcing affects evolution of water and $CO_2$ cycles and climate on Mars (e.g., Richardson and Mischna 2005).

The immediate question then is whether can one find a clear and dominant climatic impact signal by $CO_2$ radiative forcing in the glacial-interglacial transition from the current state-of-the-art modeling results from various versions of General Circulation Models (GCMs)?

---

[1] Even in this scenario of $CO_2$ emissions from using fossil-fuels, the actual amount of $CO_2$ ultimately retained in the atmosphere is still limited by climatic, chemical and biological factors of the global carbon cycle.

Let's start by imposing a $CO_2$ radiative forcing estimate of about 2.5 W/m$^2$ from the 80 ppm change which will trigger a warming of 2-3ºC (taking a high value of climate sensitivity of 1ºC per 1 W/m$^2$), and that falls far short of the calibrated 10-12ºC in Antarctic temperature change. One should note that the climate sensitivity value adopted for this simple estimate has been generous, considering the "black-body" or "no-feedback" sensitivity value of 0.3ºC per W/m$^2$ or the accepted range of values from 0.4 to 1.2ºC per W/m$^2$ after accounting for net gain from all the positive and negative feedback processes (see Joshi et al. 2003; Forster and Taylor 2006). A more concrete estimate comes from recent simulations of LGM by Schneider von Deimling et al. (2006a) where the $CO_2$ forcing contributes about 1.8ºC (or about 31%) of the total 5.7ºC cooling over the globe (see also Figure 4b in Schneider von Deimling et al. 2006b). The amount is impressive but not predominantly large. Also it is clear from the spatial pattern of change that the effects from the ice-sheet forcing (represented as an albedo effect) are clearly more extensive and variable than effects from changes in $CO_2$ forcing.

More importantly, the notion of what is forcing and what is feedback is sufficiently confused here and has led Hansen et al. (2007) to ponder that "[c]limate sensitivity when surface properties are free to change ... reveals Antarctic temperature increase of 3ºC per W/m$^2$. Global temperature change is about half that in Antarctica, so this equilibrium global climate sensitivity is 1.5ºC per W/m$^2$, double the fast-feedback (Charney) sensitivity. Is this 1.5ºC per W/m$^2$ sensitivity, rather than 0.75ºC per W/m$^2$, relevant to human-made forcings?"

The issue may not be fully resolvable for now, but it is clear that this potential double-counting of radiative "forcing" effects by $CO_2$ in Hansen et al. (2007)'s view would stand as authoritative if no contest or discussion to this problematic proposal is forthcoming. In any case, if the impacts by $CO_2$ radiative forcing were to be real for the large glacial-interglacial transitions, one should be able to verify the large stratospheric warming by up to 7ºC at 60 km predicted during the LGM by e.g., Crutzen and Bruhl (1993). It would be also important to see if the $CO_2$ forcing theory can explain the regional warming around the Seas of Japan and Okhotsk during the LGM (Ishiwatari et al. 2001; Seki et al. 2004) where it was suspected that the anomalously warm sea surface temperature may represents the local equilibrium of thermal energy from trapped solar radiation in shallow water under the highly stratified upper ocean condition of LGM when the Japan and Okhotsk seas were rather isolated from the open ocean as a result of the lowered sea level. Similarly, a correct $CO_2$'s global radiative forcing theory should also be able to account for the ice-free conditions during the LGM for coastal oasis regions of Bunger Hills (Gore et al. 2001) and Larsemann Hills (Hodgson et al. 2006) around East Antarctica.

Another way to assess the role of $CO_2$ forcing in glacial-interglacial climate change would be to study the quantitative results from variations induced by the orbital forcing alone in order to find out if there is any need to invoke $CO_2$ as a forcing input. Figure 4 shows the successful simulation of a significant snow accumulation in the glacially sensitive location (70ºN; 80ºW) around the Laurentide ice sheet area for the glacial inception scenario at orbital forcing condition around 115 kyr BP (compared to the present day orbital forcing case, PD) taking into account the coupled ocean-atmosphere feedbacks but with no change in radiative forcing by $CO_2$ (set at about 270 ppm) between 115 kyr BP and PD by Khodri et al. (2001). In other words, the Khodri et al. (2001) study confirms the important role of seasonality and the correct accounting of complex feedback mechanisms involving the atmospheric winds, ocean dynamics, and hydrologic cycles with little hint for the need of $CO_2$ forcing. These results are consistent with modeling experiments of Vettoretti and Peltier (2004). Recent glacial inception modeling experiments by Risebrobakken et al. (2007) have also essentially supported the scenario by Khodri et al. (2001) while stressing dynamics from an enhanced, rather than weakened, Atlantic meridional overturning circulation in creating a strong land-sea thermal gradient together with a strong wintertime latitudinal insolation gradient to promote increased storminess and moisture transport that feeds into the formation of the Northern European ice sheet.

Loutre and Berger (2000) further emphasized the key role of orbital solar radiation forcing in generating the glacial-interglacial cycles of ice volume changes. The authors showed that if the time-varying $CO_2$ forcing is prescribed alone, then their model is able to generate glacial-interglacial temperature changes, but unable to simulate the simultaneously varying ice volume cycles of the ice ages and interglacial warm periods. In contrast, the glacial-interglacial ice volume cycles can be generated by accounting for the orbital forcing alone with a constant level of atmospheric $CO_2$ lower than 220 ppm. As noted in section 2, such a great sensitivity of large continental ice sheet formation to the threshold-crossing at a particular low $CO_2$ level requires more in-depth scientific research, but one can offer a counter-example from other existing $CO_2$-climate modeling experiments.

Figure 5 shows the curious example of a more extensive snow accumulation over the Arctic sea area for a case of high $CO_2$ level of 290 ppm in contrast to the lower $CO_2$ level case of 260 ppm shown in Vettoretti and Peltier (2004). Both simulations were set with exact same orbital configuration of low tilt angle and high eccentricity (to emulate glacial inceptions near the terminations of MIS stages 5 and 7, respectively) so the results strictly represent the consequences of having differing levels of $CO_2$ radiative forcing. The results show more extensive snow accumulation, rather than less, with higher $CO_2$ forcing, although the rates of snow accumulation in the Canadian Arctic region and coastal eastern Siberian region are relatively higher for the low $CO_2$ case of 260 ppm. The examination of the simulated land surface temperatures and precipitation minus evaporation (P-E) anomalies in the polar region shows that although the polar surface land temperatures may be relatively warmer with less extensive cool-summer areas (i.e., latitudes with temperatures from -4 to -12ºC; see Figures 8f and 8g in Vettoretti and Peltier 2004) in the 290 ppm case, the positive P-E regions were slightly larger and enhanced in higher polar latitudes for the experiment with 290 ppm of $CO_2$ (see Figures 9f and 9g in Vettoretti and Peltier 2004). The results in Figure 5 may be consistent with the cyrospheric moisture pump scenario for glacial inceptions being relatively more effective at a higher $CO_2$ scenario, studied earlier by the same authors (Vettoretti and Peltier 2003), where an initial cooling at high latitudes by orbital insolation forcing caused evaporation to drop more quickly than precipitation locally which then created a condition favoring more moisture transport into polar regions via increased baroclinic activity at mid-to-high latitudes in the Northern Hemisphere summer season. The results of Vettoretti and Peltier (2004) is consistent with those of Khodri et al. (2001).

What about evidence for a greater role of $CO_2$ radiative forcing during deglaciation scenarios?

The recent hypothesis of Martinson and Pitman (2007) does not specifiy any prominent role for atmospheric $CO_2$ but instead details the sufficiency of a sequence of events during the last glacial termination involving the crucial role of the southward expansion of the North American and Eurasian ice sheets, sea ice cover in North Pacific and North Atlantic, and the balancing acts among Arctic freshwater budgets, salinity-driven formation of polynas (i.e., ice-free areas), and the deepwater formation in the Arctic Ocean and its subsequent overflow into the North Atlantic as well as the incursion of warm surface water from the North Atlantic to the Arctic Ocean. This hypothesis is consistent with the dynamical changes of land and sea ice contribution to the sedimentary record from central Arctic Ocean by Spielhagen et al. (1997) that also did not invoked any contribution from $CO_2$ radiative forcing.

Figure 6 shows the land ice thickness simulations of the deglaciation scenario from Yoshimori et al. (2001) for orbital conditions from 21 kyr (LGM) to 11 kyr BP (early Holocene), first with a constant $CO_2$ at 200 ppm, and then with level of $CO_2$ changed by 80 ppm for the early Holocene in order to be consistent with ice-core air bubble results. The results clearly suggest a minimal impact of added radiative forcing of $CO_2$ on thickness of ice sheet on land. Yoshimori et al. (2001) did argue for a "powerful" feedback role by $CO_2$ forcing in explaining glacial termination, but the authors pointed out that the effect of increasing atmospheric $CO_2$ from 200 to 280 ppm in their simulation leads to a nominal impact on winter air temperatures over continents adjacent to the North Atlantic. That $CO_2$ impact in turn contributes to ice-sheet nourishments through slightly enhanced winter precipitation, so $CO_2$ acts as negative, rather than positive, feedback for ice-sheet retreat during deglaciation.

The examples in Figures 4, 5 and 6 serve only as the sufficient-but-not-necessary condition of orbital insolation forcing in accounting almost fully for conditions and changes during the glacial-interglacial transition without the need to invoke the argument for $CO_2$ as the predominant amplifier of those changes.

One cannot totally discount other contemporary studies (Weaver et al. 1998; Pepin et al. 2001; Lea 2004) that suggested a "dominant" contribution by $CO_2$ radiative forcing to the observed glacial-interglacial temperature change while perhaps ignoring the large changes in global ice volume and its effects. Lea (2004) claimed that "modeling results for the glacial oceans support the hypothesis that $CO_2$ variations are the dominant source of radiative forcing in the tropical ocean regions" citing Hewitt and Mitchell (1997),

Weaver et al. (1998) and others, but simulation results from Hewitt and Mitchell (1997) estimated the lowering of $CO_2$ at 21 kyr BP at the LGM to be a cooling of 1.4ºC or about one-third of the total simulated cooling. Weaver et al. (1998) who suggested that "the most important [more so than ice-sheet albedo feedbacks] of these forcings in our model is the change in atmospheric $CO_2$" but concurrently admitted to underestimating the "ice albedo" effects. More importantly, the model of Weaver et al. (1998) suggests that temperatures in the tropics were 2.2ºC less than today's and their results show apparent insensitivity to changes in oceanic circulation. These results are not consistent with the relatively larger amplitude change in tropical SST of 3.5-4ºC during the last two glacial-interglacial transitions as deduced by Visser et al. (2003), and with the importance of oceanic feedbacks for glacial inception scenarios identified with the coupled ocean-atmosphere GCM as discussed by Khodri et al. (2001) and found in climate sensitivity experiments conducted by Vettoretti and Peltier (2004).

**5. Conclusions**

There is no quantitative evidence that varying levels of minor greenhouse gases like $CO_2$ and $CH_4$ have accounted for even as much as half of the reconstructed glacial-interglacial temperature changes or, more importantly, for the large variations in global ice volume on both land and sea over the past 650 kyr. This paper shows that changes in solar insolation at climatically sensitive latitudes and zones exceed the global radiative forcings of $CO_2$ and $CH_4$ by several-fold, and that regional responses to solar insolation forcing will decide the primary climatic feedbacks and changes (see also independent research and conclusions by Kukla and Gavin 2005; Lorenz et al. 2006; Roe 2006).

Persistent orbitally-moderated insolation forcing is, therefore, likely to be the principal driver of water vapor cycling, and the cloud-and-ice insulator and albedo feedbacks. Such a forcing-response scenario has not received enough attention in current research (but with notable exceptions, e.g., Dong and Valdes 1995; Gupta et al. 1996; Broecker 1997; Greene et al. 2002; Leduc et al. 2007). A host of other forcings and feedbacks, including dust and aerosol forcings, oceanic circulation, and vegetation cover feedbacks have not been soundly quantified. The forcing from intrinsic variation of the solar radiation and magnetic activity has also been almost entirely ignored in this paper but several recent studies are beginning to document and formulate testable climatic responses on multidecadal-to-centennual-to-millennial timescales resulting from this particularly complex expression of solar change (e.g., Bond et al. 2001; Holzkamper et al. 2004; Mayewski et al. 2004; Holzhauser et al. 2005; Maasch et al. 2005; Soon 2005; Scherer et al. 2006). There are still questions about how orbital forcings explain glaciation and deglaciation over the past few million years (Roe 2006; Tziperman et al. 2006; Huybers 2007; Lisiecki and Raymo 2007) with the 100 kyr glacial-interglacial cycles not fully explained (but most likely, nonlinearly related to the obliquity forcing emphasizing the key role of the insolation gradient as the driver for climatic processes and feedbacks, as discussed by Raymo and Nisancioglu 2003; Liu and Herbert 2004; Loutre et al. 2004; Huybers and Wunsch 2005; Huybers 2007).

However, the popular notion of $CO_2$ and $CH_4$ radiative forcing as the predominant amplifier of glacial-interglacial phase transitions cannot be confirmed. In this context, the graph of "radiative perturbation" during the last glacial maximum that is shown as the top left panel of Figure 6.5 on p. 451 of the IPCC (2007) Working Group I report, suggesting that the "global annual mean radiative influences" by orbitally-moderated insolation forcing is negligible when compared to "radiative influences" of "$CO_2$", "$CH_4 + N_2O$", "Mineral Dust", "Continental ice and sea level" and "Vegetation", may be gravely misleading. All the listed "influences" are very likely to be the responses from the initial orbital insolation forcing and its persistent effects. Provided that the deduced amplitude of 80 ppm and 400 ppb for $CO_2$ and $CH_4$ from air-bubble records is not severely underestimated, enhanced greenhouse effects from these two minor greenhouse gases cannot explain the greater part of the large climatic swings and substantial hydrologic and cryospheric changes reconstructed for the glacial-interglacial transitions over the last 650 kyr.

Our basic hypothesis is that long-term climate change is driven by solar insolation changes, from both orbital variations and intrinsic solar magnetic and luminosity variations. This implies natural warming and cooling variations on decades through millennia (e.g., Bond et al. 2001; Holzkamper et al. 2004; Holzhauser et al. 2005; Maasch et al. 2005; Soon 2005) together with an eventual cooling of the Earth and an increase in ice mass accumulation within the past-and-future horizons of the Holocene and the next few thousand years or so (see Kukla and Gavin 2005 and Bauch and Kandiano 2007 for the pioneering research and more in-depth discussion). Such a retrodiction appears consistent with proxy evidence indicating both systematic and significant cooling trends during the Holocene[2] in Greenland (Johnsen et al. 2001), western Arctic (Kaufman et al. 2004), Nordic Seas (Andersen et al. 2004), other regions around northeast Atlantic and Mediterranean (Marchal et al. 2002; Kim et al. 2007; Magny et al. 2007), northwest Atlantic (Sachs 2007), northwest Africa and Gulf of Guinea (Kim et al. 2007; Weldeab et al. 2007), western tropical Pacific ocean (Stott et al. 2004), southern midlatitudes at seas around the Indian Ocean and the Australian-New Zealand region (Ikehara et al. 1997; Pahnke and Sachs 2006; Barrows et al. 2007) and even the Antarctic Peninsula and both coastal and inland region of East Antarctica (Masson et al. 2000; Masson-Delmotte et al. 2004; Hodgson et al. 2006; Smith et al. 2007). This predictable tendency and currently observed reality led Sachs (2007) to the conclusion, with which we agree, that "the Holocene, which often considered a time of climate stability, was characterized by large secular changes throughout the climate system. Perhaps [this cooling] is a harbinger of climate deterioration preceding the next glacial period."

**Acknowledgements** – I thank Guido Vettoretti and Thomas Schneider von Deimling for professional courtesy in answering some questions concerning their papers, Masakazu Yoshimori for sharing his figures and PhD thesis, and especially Eugene Avrett and Christopher Monckton for significant improvements in several drafts. I further thank the three referees, especially referee C, for constructive reviews. Than, Lien, and Julia Pham, and Benjamin and Franklin Soon are acknowledged for loving support and motivation.

---

[2] Our brief discussion on the Holocene changes is not intended to be complete, see additional discussion in Mayewski et al. (2004), Lorenz et al. (2006) and the Holocene SST database offered by GHOST (Global Holocene Spatial and Temporal Climate Variability) at http://www.pangaea.de/Projects/GHOST.

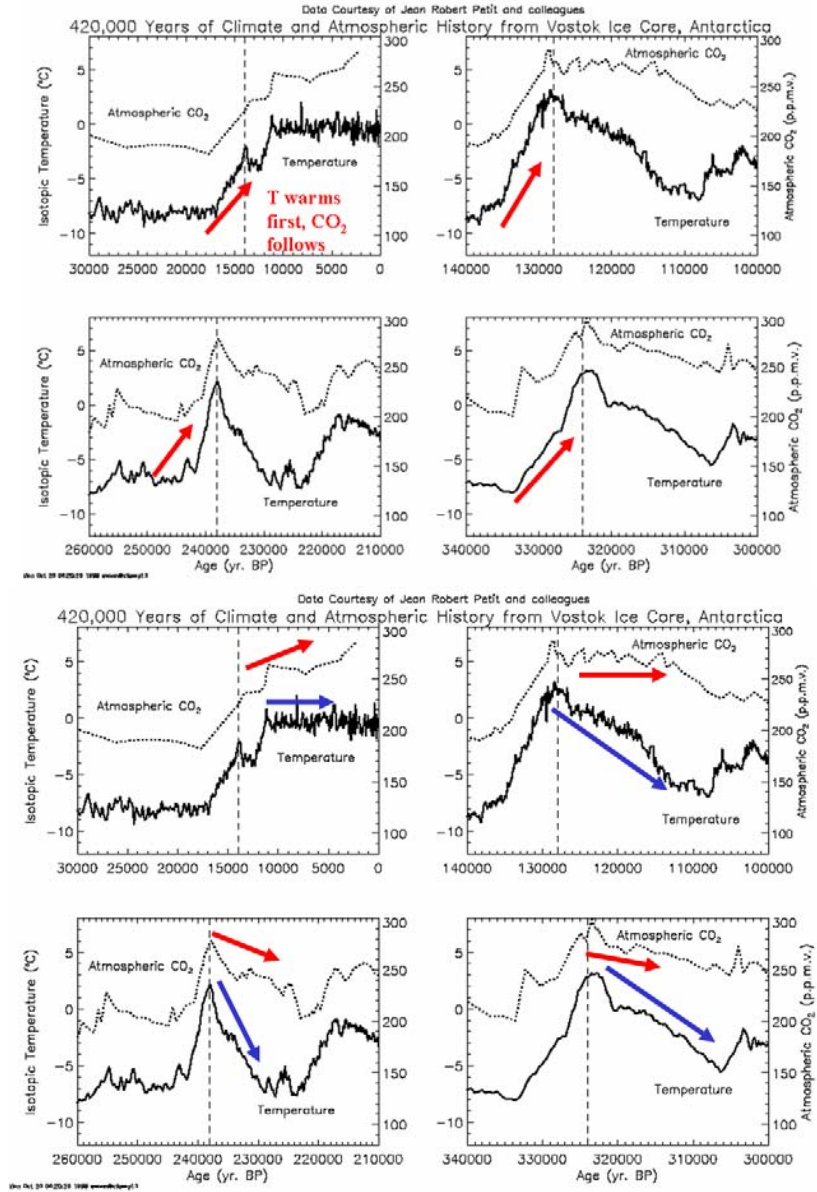

**Figure 1:** Vostok temperature and atmospheric $CO_2$ history for the past 420 kyr, from Petit et al. (1999), showing that the Antarctic warming tends to lead the rise in $CO_2$ concentrations by several hundred years during the last four deglaciations (upper panel) and that relatively high $CO_2$ levels can be sustained for thousand of years during glacial inception scenarios when the temperature has dropped significantly (lower panel) [see Fischer et al. 1999]. Cuffey and Vimeux (2001) and Vimeux et al. (2002) showed that the co-variation of the Vostok atmospheric $CO_2$ and isotopic temperature, once corrected for effects from changes in moisture source for the temperature, came to much closer timing agreement for the last 150 kyr BP but as discussed in Section 1, the atmospheric $CO_2$ content must be still be somehow controlled by other local and regional climatic variables.

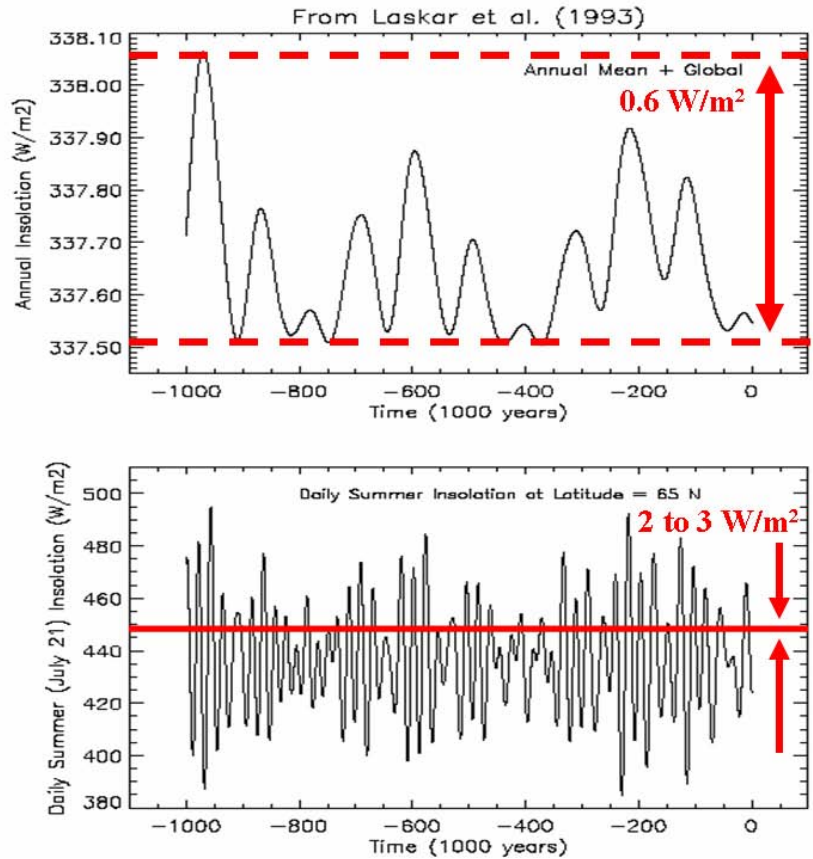

**Figure 2:** Sensitivity of the Earth climate to incoming radiation from the Sun over the past million years, based on calculations by Laskar et al. (1993) accounting (only) for the geometrical changes in the Sun-Earth orbit. First, the climate receives and reacts to local midsummer insolation (here taken at July 21 at 65ºN: lower panel). The influence of persistent, daily, localized insolation at midsummer values is demonstrably of very much greater climatic effect, and hence more relevant in assessing the contribution of insolation to the paleoclimate, than the global annual mean insolation (upper panel) or the paleoclimatological forcing from $CO_2$ (of about 2 to 3 $W/m^2$; Joos 2005). (For a more direct comparison of solar insolation quantity presented here to, say, the radiative forcing from atmospheric $CO_2$, one needs to weigh in the additional effect of reflection of sunlight by the Earth system).

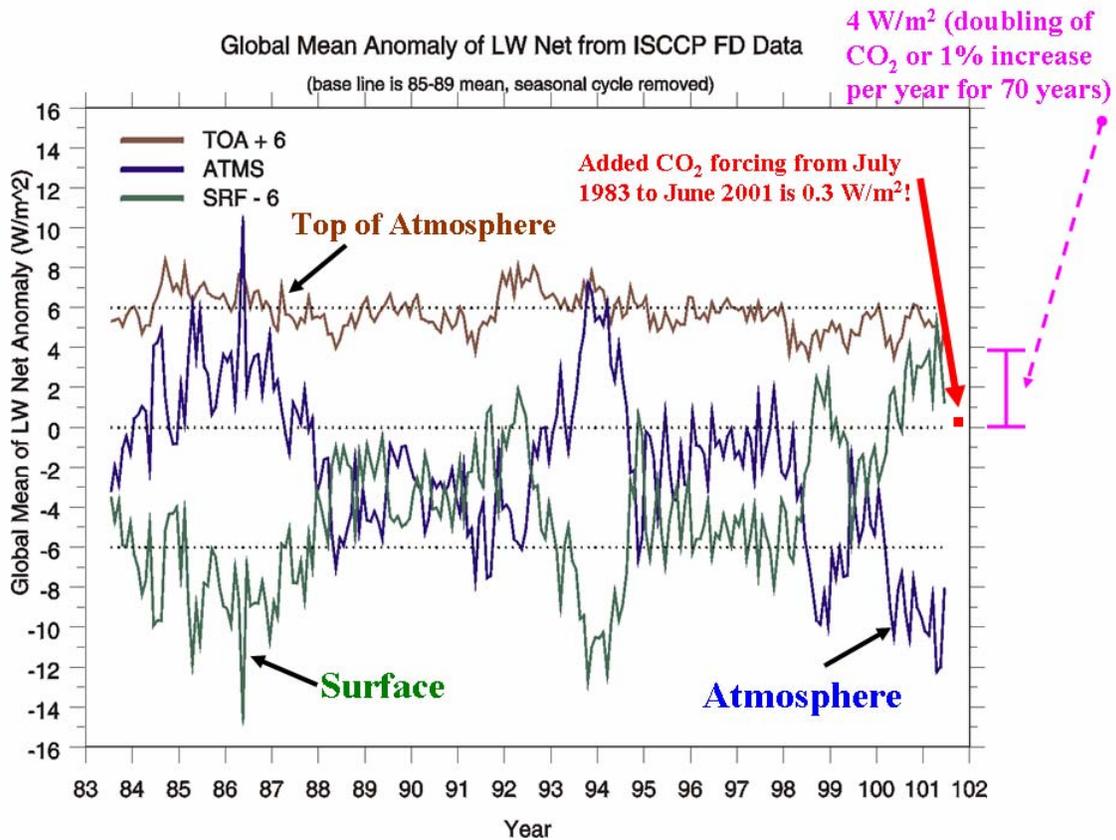

**Figure 3:** Global net longwave (LW) fluxes deduced for the surface, the atmospheric column and at the top of the atmosphere from July 1983 through June 2001 in comparison to the estimated radiative forcing of 0.3 W/m$^2$ from increased anthropogenic $CO_2$ over the same 18-year time-span as well as the 4 W/m$^2$ estimate for a doubling of atmospheric $CO_2$ which roughly extends over a 70-year period if one compounded the $CO_2$ concentration increase at a rate of 1% per year [adapted from the original figure shown in International Satellite Cloud Climatology Project, ISCCP, web page http://isccp.giss.nasa.gov/projects/flux.html with technical discussion in Zhang et al. (2004)]. Recent and projected $CO_2$ forcing are likely to be confused by the LW variations produced by internal variability and solar radiation-induced forcing and feedback through water vapor and cloud variations.

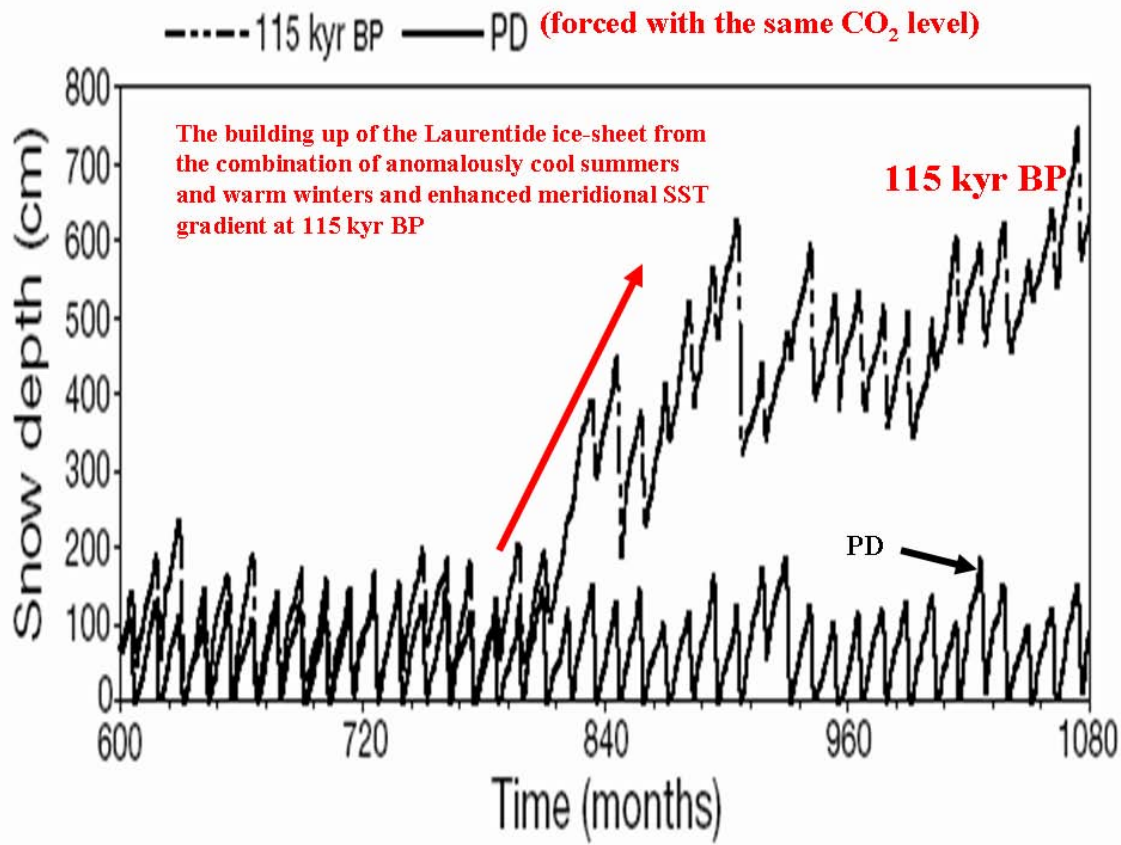

**Figure 4:** The successful simulations of a significant snow accumulation in the glacially sensitive location (70ºN; 80ºW) around the Laurentide ice sheet area for the glacial inception scenario with orbital forcing condition around 115 kyr BP (compared to the present day orbital forcing case, PD) taking into account the coupled ocean-atmosphere feedbacks but with no change in $CO_2$ forcing (set at about 270 ppm) between 115 kyr BP and PD [adapted from Khodri et al. (2001)]. The Laurentide ice sheet was probably formed without much help from $CO_2$ forcing.

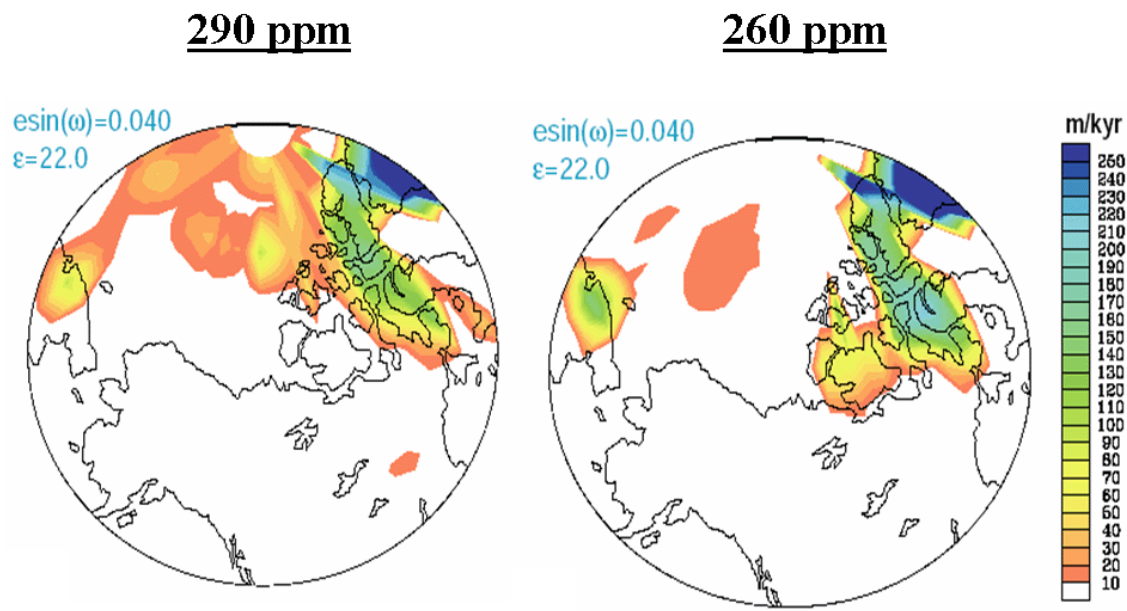

**Figure 5:** Perennial snow accumulation rate (in m/kyr) for low tilt angle ($\varepsilon = 22.0°$) and highly eccentric (e $\sin(\omega) = 0.04$ with precession parameter $\omega = 90°$) orbital (i.e., near the glacial inception phase) scenarios for $CO_2$ levels set at 290 ppm (left panel) [roughly corresponds to the termination of marine isotope stage 5; 115 kyr BP] and 260 ppm (right panel) [roughly corresponds to the termination of marine isotope stage 7; 220 kyr BP] [adapted from Vettoretti and Peltier (2004)]. Greater $CO_2$ forcing yields larger snow accumulation over the Arctic Ocean (see discussion in the text).

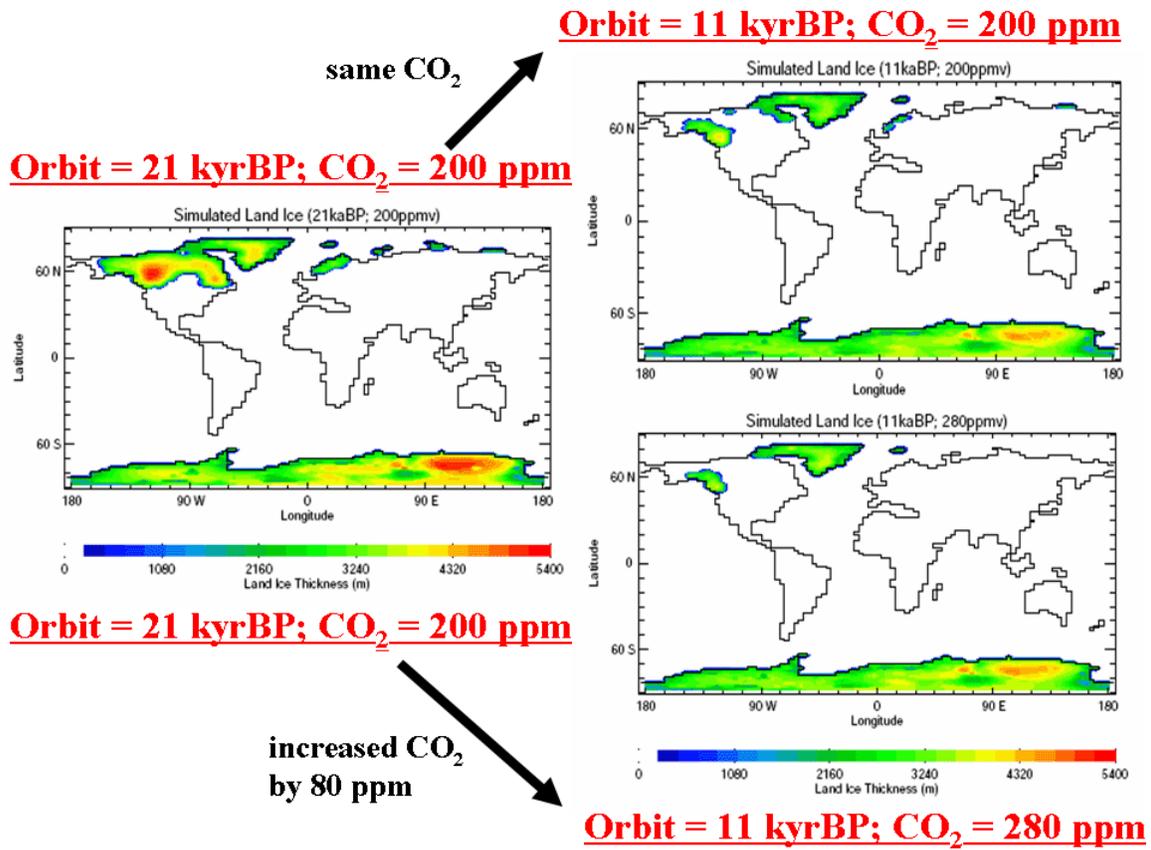

**Figure 6:** Simulations of land-ice thickness (in m) for orbital deglaciation scenarios between LGM (21 kyr BP) and early Holocene (11 kyr BP) with the same $CO_2$ level at 200 ppm (right panel →top left panel) and $CO_2$ level increased from 200 ppm at LGM to 280 ppm at early Holocene (right panel →bottom left panel) [adapted from Yoshimori et al. (2001)]. Orbitally-induced solar forcing causes greater land-ice response than even a significant increase in atmospheric $CO_2$ concentration by 80 ppm.